\documentclass[twocolumn,superscriptaddress,showpacs]{revtex4}
\usepackage{calligra}
\usepackage{amssymb}
\usepackage{amsmath}
\usepackage{txfonts}
\usepackage{graphicx,latexsym}
\usepackage{subfigure}
\DeclareMathOperator{\Tr}{Tr}\DeclareMathOperator{\C}{C}\DeclareMathOperator{\N}{N}

\DeclareMathOperator{\RIBQDV}{RIBQDV} \DeclareMathOperator{\qd}{QD}
\DeclareMathOperator{\PVNE}{PVNE} \DeclareMathOperator{\REE}{REE}
\DeclareMathOperator{\EOF}{EOF} \DeclareMathOperator{\OWQD}{OWQD}
\DeclareMathOperator{\qc}{QC} \DeclareMathOperator{\RIMPED}{RIMPED}
\DeclareMathOperator{\RIBLMIMDV}{RIBLMIMDV}\DeclareMathOperator{\GQD}{GQD}\DeclareMathOperator{\hsd}{HSD}
\DeclareMathOperator{\RIBHSDV}{RIBHSDV}\DeclareMathOperator{\HSD}{HSD}
\DeclareMathOperator{\LMIMD}{LMIMD}
\DeclareMathOperator{\LEMID}{LEMID}
\DeclareMathOperator{\LEMIMD}{LEMIMD}
\DeclareMathOperator{\GMLMIMD}{GMLMIMD}
\DeclareMathOperator{\GMLEMID}{GMLEMID}
\DeclareMathOperator{\GMLEMIMD}{GMLEMIMD}
\DeclareMathOperator{\LMLMIMD}{LMLMIMD}

\DeclareMathOperator{\LMLEMID}{LMLEMID}
 \DeclareMathOperator{\ggqd}{GGQD}
\DeclareMathOperator{\gqd}{GQD} \DeclareMathOperator{\gmgqd}{GMGQD}
 \DeclareMathOperator{\LMQD}{LMQD}
\DeclareMathOperator{\LMGQD}{LMGQD}
\DeclareMathOperator{\LMVNE}{LMVNE}
\DeclareMathOperator{\cgqd}{CGQD}
\DeclareMathOperator{\ggmGQD}{GGMGQD}
\DeclareMathOperator{\ggmLMIMD}{GGMLMIMD}
\DeclareMathOperator{\bqd}{BQD}

\DeclareMathOperator{\CQD}{CQD}
\DeclareMathOperator{\CHSD}{CHSD}\DeclareMathOperator{\CD}{CD}
\DeclareMathOperator{\GCQD}{GCQD}
\DeclareMathOperator{\GCHSD}{GCHSD}\DeclareMathOperator{\GCD}{GCD}

\DeclareMathOperator{\gmvne}{GMVNE}
\DeclareMathOperator{\gsmqd}{GSMQD} 
\DeclareMathOperator{\TD}{TD} \DeclareMathOperator{\RIBTDV}{RIBTDV}
\DeclareMathOperator{\gsmtd}{GSMTD}
\DeclareMathOperator{\gsmhsd}{GSMHSD}

\DeclareMathOperator{\tc}{TC} \DeclareMathOperator{\cc}{CC}

\usepackage{CJK}
\usepackage{amssymb}
\usepackage{graphicx}
\usepackage{epsf}
\usepackage{dcolumn}
\usepackage{bm}
\usepackage{longtable}
\usepackage{CJK}
\usepackage{color}
\begin{document}
\begin{CJK}{GBK}{song}

\title{Entanglement as Difference: Reduction-induced Minimal Partial Entropy Difference}
\author{Jing-Min Zhu} \email{zhjm-6@163.com}
\affiliation {College of Optoelectronics Engineering (College of
Weather Modification), Chengdu University of Information Technology,
Chengdu 610225, China}

\begin{abstract}

Bipartite mixed-state quantum entanglement (QE) and its measures
play a crucial role in both theoretical research and practical
quantum applications. Its internal structure is far more complex and
less well understood compared with bipartite pure-state QE. Some
existing measures involve inherently intractable global optimizations, 
while others are only applicable to highly
limited-dimensional quantum systems.
Here based on the inherent feature that bipartite QE systems
nonseparable necessarily implies that local reduced density matrix
differs from its \textquotedblleft native\textquotedblright density matrix, 
we propose a more physical and intuitive measure termed
Reduction-induced Minimal Partial Entropy Difference to quantify
arbitrary bipartite mixed-state QE. 
Partial Von Neumann Entropy is only a pure-state special
case of this method.
This measure offers intrinsic structural 
insights into bipartite QE characterization, thereby establishing
itself as a valuable complementary measure. Its intuitive and clear
physical
picture, combined with wide applicability 
and relatively low computational complexity, facilitates exploring
its potential quantum information applications, hence its conceptual
framework and line of thought deserve to be further developed to
describe and quantify multipartite QE in the future.

\vspace{0.1in} \noindent{\bf \small{Keywords}}:Quantum Entanglement
(QE); Bipartite Mixed-state QE; Reduction-induced Minimal Partial
Entropy Difference



\end{abstract}
 \makeatletter
\makeatother

\maketitle
\section{Introduction}


In 1935, \textquotedblleft Spooky Action at a
Distance\textquotedblright described as a strange phenomenon of
quantum mechanics challenged the completeness of the quantum theory
by the famous physicists Einstein, Podolsky and Rosen (EPR)
\cite{epr}. And then in the same year, the concept of quantum
entanglement (QE) was first proposed clearly by Schr\"{o}dinger
\cite{sd}, through
rigorous mathematical formalism, he 
demonstrated that entangled states cannot be expressed as a simple
product of the states of their individual constituents; rather, they
must be treated as an inseparable composite system. Subsequently, in
1964, Bell \cite{bell} showed that quantum entanglement leads to
experimentally testable predictions that deviate from classical
physics. Despite decades of research, the fundamental nature of
quantum entanglement continues to challenge and deepen our
understanding of the laws governing the physical world.

The type or degree of QEs are often directly related to their
revolutionary applications including quantum cryptography
\cite{kk,ng}, quantum teleportation \cite{ch,rc,nr}, quantum
superdense coding \cite{nr,sj}, measurement-based quantum
computation \cite{rr,fa} and quantum enhanced sensing
\cite{tc,rcp,lp}. Therefore, a deeper understanding and accurate
metric of QE have become important prerequisites for the development
of quantum information science and quantum technology. However, to
date, the quantum mechanical structure and measure of QE in
bipartite mixed states remains unclear; let alone the organization
and structure, description and measure of multipartite QEs. Here,
our primary focus is on QEs in bipartite mixed states.

Mixed states are more reflective of real-world scenarios, as factors
like environmental conditions, quantum measurements and thermal
influences can lead quantum systems to exist in mixed states.
Therefore, the quantification of QE in bipartite mixed states is
more meaningful in practice, aligning more effectively with
experimental conditions and practical applications. Consequently,
the study of bipartite mixed-state QE and its quantification holds
considerable significance not only from a theoretical
perspective\text{---}rooted in the fundamental principles of quantum
mechanics\text{---}but also in practical research and technological
implementations.


\section{existing Bipartite QE measures}
\subsection{Partial Von Neumann Entropy for Bipartite Pure-state QE}
Nonseparable bipartite systems exhibit QE \cite{epr,sd}, which can
be quantified using various measures. For a pure bipartite state
$\mid\Psi\rangle$, its density matrix
$\rho=\mid\Psi\rangle\langle\Psi\mid$, a well-known and commonly
used bipartite pure-state QE measure, the Partial Von Neumann
Entropy (PVNE) \cite{ch1,ch2} is defined as
\begin{eqnarray}
\PVNE(\rho)=S(\rho_i)=-\Tr (\rho_i \log_{2}{\rho_i}),
\end{eqnarray}
where $\rho_i (i=1,2)$ represents local reduced density matrix and
the symbol $\Tr () $ denotes taking the trace of a density matrix,
the sum of the elements on its main diagonal. This Partial Von
Neumann Entropy can effectively measure the degree of QE between
bipartite pure states, while for bipartite mixed-state QE, it is
ineffective.

\subsection{Entanglement of Formation and Concurrence}

For bipartite QE, especially for bipartite mixed-state QE, an
extensively acknowledged type of measure, Entanglement of Formation
(EOF) \cite{ch1,ch2} was introduced and is defined as
\begin{eqnarray}
\EOF(\rho)=\min_{\{p_j,\mid\Psi_j\rangle\}}\sum_j
p_j\PVNE(|\Psi_j\rangle\langle\Psi_j|),
\end{eqnarray}
which represents the minimum average value over all possible
ensemble pure-state decompositions of the bipartite density matrix
$\rho =\sum_j p_j\mid\Psi_j\rangle\langle\Psi_j\mid$, where $p_j>0$
and $\sum_j p_j=1$. For bipartite pure-state QE, the Entanglement of
Formation reduces to the Partial Von Neumann Entropy
\begin{eqnarray}
\EOF(\mid\Psi\rangle)=\PVNE(\mid\Psi\rangle)=S(\rho_i)=-\Tr (\rho_i
\log_{2}{\rho_i}).
\end{eqnarray}
However, for mixed-state QE, in general, this measure heavily relies
on the various decompositions of pure states, and to date, no universal algorithm exists to determine the minimum QE decomposition. 
To address this issue and improve the Entanglement of Formation
behavior in mixed states, Reference \cite{wk} discovered that for
two-qubit mixed states, the Entanglement of Formation reduces to
\begin{eqnarray}
\EOF(\rho)=H(z)=-z \log_{2}(z)-(1 - z) \log_{2}(1-z),
\end{eqnarray}
where the argument z is given by
\begin{eqnarray}
z = \frac{1 + \sqrt{1 - C^2}}{2}, 
\end{eqnarray}
and $C(\rho)$ denotes Concurrence a fundamental entanglement measure
for two-qubit systems, introduced by Wootters as the analytic
solution to the entanglement of formation, which is defined as
\cite{wk,au,ka,mf,xjw}
\begin{eqnarray}
\C(\rho) = \max\{ 0, \lambda_1 - \lambda_2 - \lambda_3 - \lambda_4
\},
\end{eqnarray}
where $\{ \lambda_1, \lambda_2, \lambda_3, \lambda_4 \}$ represent
the square roots of the eigenvalues of the matrix $\rho
\tilde{\rho}$ arranged in decreasing order. The time-reversed matrix
$\tilde{\rho}$ is defined as
\begin{eqnarray}
\tilde{\rho}= (\sigma_y \otimes \sigma_y) \rho^* (\sigma_y \otimes
\sigma_y),
\end{eqnarray}
where $\sigma_y$ is the Pauli matrix and $\rho^*$ denotes the
complex conjugate of $\rho$. It is important to note that for
two-qubit states, the Concurrence serves as a widely accepted and
practically effective measure of QE. While for a bipartite
higher-dimensional pure state $\psi$, the Concurrence
\cite{pr,kc,cs,cj,zb} is defined as

\begin{eqnarray}
\C(\psi) =\sqrt{2(1-\Tr({\rho^2_1)})},
\end{eqnarray}
where the dimension of the subsystem satisfies $n_1 \leq n_2$. In
the case of a bipartite higher-dimensional mixed state $\rho$, the
Concurrence \cite{pr,kc,cs,cj,zb} is defined as follows:
\begin{eqnarray}
\C(\rho)=\min_{\{p_j,|\psi_j\rangle\langle\psi_j|\}}\sum_j
p_j\C(|\psi_j\rangle\langle\psi_j|),
\end{eqnarray}
where the minimum average value is taken over all possible ensemble
pure-state decompositions of $\rho =\sum_j
p_j|\psi_j\rangle\langle\psi_j|$, with $p_j>0$ and $\sum_j p_j=1$.
Consequently, for bipartite higher-dimensional systems, there is no
unified and straightforward formula analogous to that for two-qubit
systems. The computations of Concurrence become significantly more
complex for higher-dimensional mixed states,
similar to the complexities encountered with the Entanglement of Formation. 

\subsection{Relative Entropy of Entanglement}
Another well-known and commonly used bipartite QE measure,
especially for bipartite mixed-state QE, Relative Entropy of
Entanglement (REE) \cite{vv1,vv2} is defined as
\begin{eqnarray}
\REE=S(\rho\,\big\|\,\sigma)=\min_{\sigma \in \mathcal{S}}[\Tr [\rho
(\log_2 \rho -\log_{2}{\sigma})]],
\end{eqnarray}
where $\mathcal{S}$ represents a set of all separable bipartite
density matrix. The Relative Entropy of Entanglement measures the
shortest \textquotedblleft distance\textquotedblright from a given
quantum entangled state to the set of all separable states, which is
the geometric interpretation of this method and reflects that the
involving computation is inherently complex. While for bipartite
pure QE states, the Relative Entropy of Entanglement simplifies to
the Partial Von Neumann Entropy
\begin{eqnarray}
\REE&&=S(\rho\,\big\|\,\sigma)=\min_{\sigma \in \mathcal{S}}[-\Tr
(\rho \log_{2}{\sigma})]\nonumber\\&&=\PVNE(\rho)=S(\rho_i)=-\Tr
(\rho_i \log_{2}{\rho_i}).
\end{eqnarray}

\subsection{Negativity for $2\times2$, $3\times2$ and $2\times3$ quantum
systems}

Negativity is a computable measure of QE based on the positive
partial transpose (PPT) criterion \cite{K.Z,gv}. For a bipartite
mixed state $\rho$, its partial transpose with respect to subsystem
i, denoted $\rho^{T_i}$, is obtained by transposing the indices of
subsystem i while leaving the other subsystem unchanged. The PPT
criterion states that if $\rho$ is separable, then $\rho^{T_i}$ must
be positive semi-definite, i.e., all its eigenvalues are
non-negative. Consequently, the presence of any negative eigenvalue
$\lambda_{\min}$ in $\rho^{T_i}$ signals the existence of
entanglement. The Negativity $N(\rho)$ is defined as twice the
absolute value of this Negative eigenvalue \cite{K.Z,gv}.
Mathematically, it is expressed as:
\begin{equation}
\N(\rho)=2 \max(0,-\lambda_{\min}).
\end{equation}
For $2\times2$, $3\times2$ and $2\times3$ systems, the PPT criterion
is a necessary and sufficient condition for separability, rendering
Negativity a complete entanglement measure. However, in
higher-dimensional systems, there exist bound entangled states that
are PPT yet inseparable; for such states, Negativity vanishes
identically, failing to detect the underlying entanglement.

QE in bipartite mixed states is inherently more complex and less
well understood than in bipartite pure states, which are generally
more intuitive and straightforward to analyze. 
The established five bipartite QE measures mentioned above exhibit
various limitations. For instance, Partial Von Neumann Entropy is
restricted to pure bipartite quantum systems; Entanglement of
Formation and Relative Entropy of Entanglement involve inherently
intractable global optimizations, as does Concurrence when applied
to higher-dimensional mixed states. Meanwhile
Negativity is limited to quantum systems of dimensions $2\times2$,
$3\times2$ and $2\times3$. 
In this context, our research aims to explore and develop new
metrics for quantifying bipartite mixed-state QE in detail, as
outlined below.



\section{Reduction-induced Minimal Partial Entropy Difference}
In any quantum system, all the information is contained
within its wave function or density matrix $\rho$. 
A bipartite QE state is inherently nonseparable. When a subsystem is
obtained through partial tracing or averaging out the freedom
degrees of the other subsystem, some information is inevitably lost.
Consequently, the local reduced density matrix of a single party,
$\rho_i$ $(i=1,2)$, differs from its \textquotedblleft
native\textquotedblright density matrix denoted by
$\rho_i^{\mathrm{nat}}$, which encodes the full local structural
information of that party as derived from $\rho$. Therefore the
concept of \textquotedblleft Entanglement as
Difference\textquotedblright is presented.



Any bipartite mixed entangled state $\rho$ can be decomposed, based
on its eigenvalues and eigenvectors, into a convex combination of
entangled pure states with the minimal possible number and a
maximally separable component \cite{ls}, which optimally
characterizes its QE, as expressed below
\begin{eqnarray}
\rho= \sum_{k=1}^{N_{\min}} p_k |\psi_k\rangle\langle\psi_k|
+\sigma^{\text{max}}_{\text{sep}},
\end{eqnarray}
where $|\psi_k\rangle$ represents entangled pure state occurring
with probability $ p_k$, $\sigma^{\text{max}}_{\text{sep}}$ is a
maximally separable component, and $N_{\min}$ is the minimal number
of entangled pure states required for such a decomposition. 
Dual to maximizing the weight of the separable component
$\sigma^{\text{max}}_{\text{sep}}$ within the decomposition, we
propose a more physical and intuitive measure\text{---}Entanglement
as Difference: Reduction-induced Minimal Partial Entropy
Difference\text{---}defined in the Schmidt basis of the entangled pure-state subsystems, 
to quantify QE in
arbitrary bipartite mixed states, as defined below 
\begin{eqnarray}
\RIMPED(\rho) = \min_{\text{decompositions}} \; S\bigl(\rho_i
\bigr)- S\bigl(\rho_i^{\mathrm{nat}}\bigr).
\end{eqnarray}
This measure is inherently subsystem-dependent. It vanishes if and
only if the state is separable, and on average, does not experience
an increase under Stochastic Local Operations and Classical
Communications, thereby establishing itself as an entanglement
monotone \cite{G.V,Y.C,W.J}.
Notably, this measure by directly extracting the minimal entangled
component and the maximal separable component based on the system's
eigenvalues and eigenvectors, 
achieves lower computational complexity than the intractable global
optimizations required by Entanglement of Formation, Relative
entropy of Entanglement and Concurrence for higher-dimensional mixed
states, providing a
practical balance between computational efficiency and generality. 
Its intuitive and clear physical picture, along with a wide range of
applications and comparatively modest computational requirements,
facilitates the investigation of its specific potential applications
in quantum information science. In the case of any bipartite
pure-state QE, this measure reduces to the Partial Von Neumann
Entropy
\begin{eqnarray}
\RIMPED(\rho)=\PVNE(\rho)=S(\rho_i)=-\Tr (\rho_i \log_{2}{\rho_i}).
\end{eqnarray}
Partial Von Neumann Entropy is only a pure-state special case of
this method, 
confirming the theoretical generality of this method. This measure offers intrinsic structural 
insights into bipartite QE characterization, establishing itself as
a valuable complementary measure.
Subsequently, to evaluate its behavior and position within the field
of existing bipartite QE measures, we will perform a comprehensive
comparative analysis across three examples perfectly covering
various structural complexities from pure Bell-like states to
directional structured mixed states and symmetric global mixed
Werner states, as detailed below.


\section{Its application and comparative analysis}
We initially consider the family of symmetric pure Bell-like states
in a noiseless environment
\begin{equation}
|\Psi\rangle=\cos\alpha|11\rangle \pm \sin\alpha|00\rangle
\end{equation}
and
\begin{equation}
|\Phi\rangle=\cos\alpha|10\rangle \pm \sin\alpha|01\rangle
\end{equation}
with $0\leqslant\alpha\leqslant\frac{\pi}2$.
Consequently, the Reduction-induced Minimal Partial Entropy
Difference, the Entanglement of Formation and the Relative Entropy
of Entanglement all consistently simplify to the Partial Von Neumann
Entropy:
\begin{eqnarray}
&&\RIMPED(\rho)=\PVNE(\rho)=S(\rho_i)=\EOF(\rho)=\REE(\rho)\nonumber\\&&=-(\cos^2\alpha\log_{2}{\cos^2\alpha}+\sin^2\alpha\log_{2}{\sin^2\alpha}).
\end{eqnarray}
Additionally, the Concurrence and Negativity consistently can be
expressed as
\begin{equation}
\C(\rho)=\N(\rho)=|\sin{2\alpha|}.
\end{equation}
This highlights the interference properties and indicates a clear
and intuitive relationship between QE and quantum coherence. The
previously mentioned measures are illustrated in Fig.1 which shows
that when $\alpha=\frac \pi4$ corresponding to the maximally
entangled Bell state, all the measures satisfy:
$\RIMPED(\rho)=\PVNE(\rho)=S(\rho_i)=\EOF(\rho)=\REE(\rho)=\C(\rho)=\N(\rho)=1$;
whereas for $\alpha=0, \frac{\pi}2$,
$\RIMPED(\rho)=\PVNE(\rho)=S(\rho_i)=\EOF(\rho)=\REE(\rho)=\C(\rho)=\N(\rho)=0$;
and for the ranges $0<\alpha<\frac\pi 2$ and $\frac\pi
4<\alpha<\frac\pi 2$, the relationships
$\RIMPED(\rho)=\PVNE(\rho)=S(\rho_i)=\EOF(\rho)=\REE(\rho)<\C(\rho)=\N(\rho)$
hold. Clearly, for the specified pure-state QEs, 
the monotonicities of these measures are fully consistent. The
Reduction-induced Minimal Partial Entropy Difference, Entanglement
of Formation and Relative Entropy of Entanglement all reduce to the
Partial Von Neumann Entropy, but is less than or equal to
Concurrence and Negativity  here signifying quantum coherence.







\begin{figure}
\setlength{\abovecaptionskip}{0.1pt}%
\setlength{\belowcaptionskip}{0.1pt}%
\begin{center}
\includegraphics[width=0.45\textwidth]{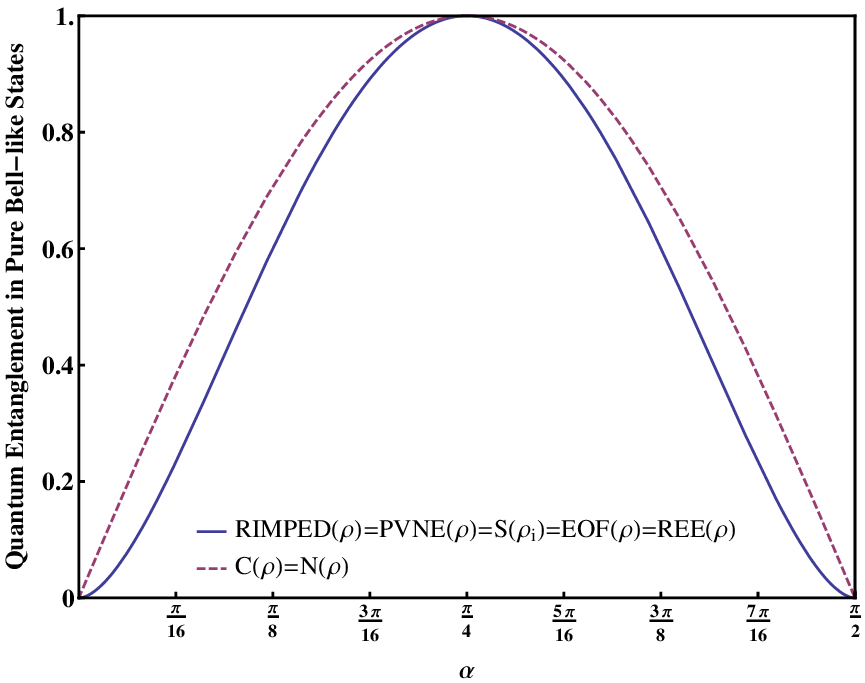}
\renewcommand{\figurename}{Fig.}
\caption{
Reduction-induced Minimal Partial Entropy Difference as well as
other some established QE measures in pure Bell-like states varying
with $\alpha$. When $\alpha=\frac \pi4$,
$\RIMPED(\rho)=\PVNE(\rho)=S(\rho_i)=\EOF(\rho)=\REE(\rho)=\C(\rho)=\N(\rho)=1$;
whereas for the values $\alpha=0, \frac{\pi}2$,
$\RIMPED(\rho)=\PVNE(\rho)=S(\rho_i)=\EOF(\rho)=\REE(\rho)=\C(\rho)=\N(\rho)=0$;
and in the ranges $0<\alpha<\frac\pi 4$ and $\frac\pi
4<\alpha<\frac\pi 2$,
$\RIMPED(\rho)=\PVNE(\rho)=S(\rho_i)=\EOF(\rho)=\REE(\rho)<\C(\rho)=\N(\rho)$.}
\end{center}
\end{figure}


We next consider the bipartite symmetric directional structured
mixed states under directional noise
\begin{eqnarray}
&&\rho=\frac {\sin^2\alpha} 2 \mid10 \pm 01\rangle\langle10 \pm 01|
+ \cos^2\alpha|00\rangle\langle00\mid
\end{eqnarray}
with $0\leqslant\alpha\leqslant\pi$, inherently realizing the Reduction-induced Minimal achievable Partial Entropy Difference. 
The first term corresponds to a entangled Bell-state with a
probability of $\sin^2\alpha$, while the second term represents a
separable pure state with probability $\cos^2\alpha$. 
The Reduction-induced Minimal Partial Entropy Difference (RIMPED) is
computed as
\begin{eqnarray}
&\RIMPED
=-(\frac {3+\cos{2\alpha}} 4 \log_{2}{\frac {3+\cos{2\alpha}}
4}+\frac{\sin^2\alpha} 2 \log_{2}{\frac {\sin^2\alpha} 2})\nonumber\\
&+\frac{2-\sqrt{3+\cos{4\alpha}}}
4\log_{2}{\frac{2-\sqrt{3+\cos{4\alpha}}}
4}\nonumber\\
&+\frac{2+\sqrt{3+\cos{4\alpha}}}
4\log_{2}{\frac{2+\sqrt{3+\cos{4\alpha}}} 4},
\end{eqnarray}
in addition to other QE measures like the Entanglement of Formation
given by
\begin{eqnarray}
\EOF(\rho)=&&-\frac {1-\sqrt{1-\sin^4\alpha}} 2 \log_{2}{\frac
{1-\sqrt{1-\sin^4\alpha}} 2}\nonumber\\
&&-\frac {1+\sqrt{1-\sin^4\alpha}} 2 \log_{2}{\frac
{1+\sqrt{1-\sin^4\alpha}} 2},
\end{eqnarray}
the Negativity expressed as
\begin{eqnarray}
\N(\rho)=\sqrt{\frac {1+\cos^2{2\alpha}} 2}-\cos^2{\alpha},
\end{eqnarray}
and the Relative Entropy of Entanglement as well as Concurrence
given by
\begin{eqnarray}
\REE(\rho)=\C(\rho)=\sin^2\alpha,
\end{eqnarray} which represents the probability for the system in the entangled Bell-state. 
The aforementioned five bipartite QE measures are depicted in Fig.2
which shows that when $\alpha=\frac \pi 2$ corresponding to the
maximally entangled Bell state, the following relations hold:
$\RIMPED(\rho)=\EOF(\rho)=\REE(\rho)=\C(\rho)=\N(\rho)=1$; whereas
for $\alpha=0, \pi$,
$\RIMPED(\rho)=\EOF(\rho)=\REE(\rho)=\C(\rho)=\N(\rho)=0$. For
values $\alpha=0.8, 2.341$, the relationships
$\RIMPED(\rho)=\N(\rho)=0.222676<\EOF(\rho)=0.371412<\REE(\rho)=\C(\rho)=0.515241$
hold; furthermore, in the ranges $0<\alpha<0.8$ and
$2.341<\alpha<\pi$, we find
$\N(\rho)<\RIMPED(\rho)<\EOF(\rho)<\REE(\rho)=\C(\rho)$; in
contrast, while within the ranges $0.8<\alpha<\frac\pi 2$ and
$\frac\pi 2<\alpha<2.341$, the relations become
$\RIMPED(\rho)<\EOF(\rho)<\REE(\rho)=\C(\rho)=\N(\rho)$.
The Relative Entropy of Entanglement and Concurrence represent
the upper limit of QE, among all considered these measures, and 
reflect that they may not accurately reflect the degree of bipartite
mixed-state QE. Conversely, the Reduction-induced Minimal Partial
Entropy Difference and the Negativity are the smallest, with the
Entanglement of Formation falling in between. These differences in metric values, 
arise from their distinct physical implications and mathematical
structures. 

\begin{figure}
\setlength{\abovecaptionskip}{0.3pt}%
\setlength{\belowcaptionskip}{0.1pt}%
\begin{center}
\includegraphics[width=0.45\textwidth]{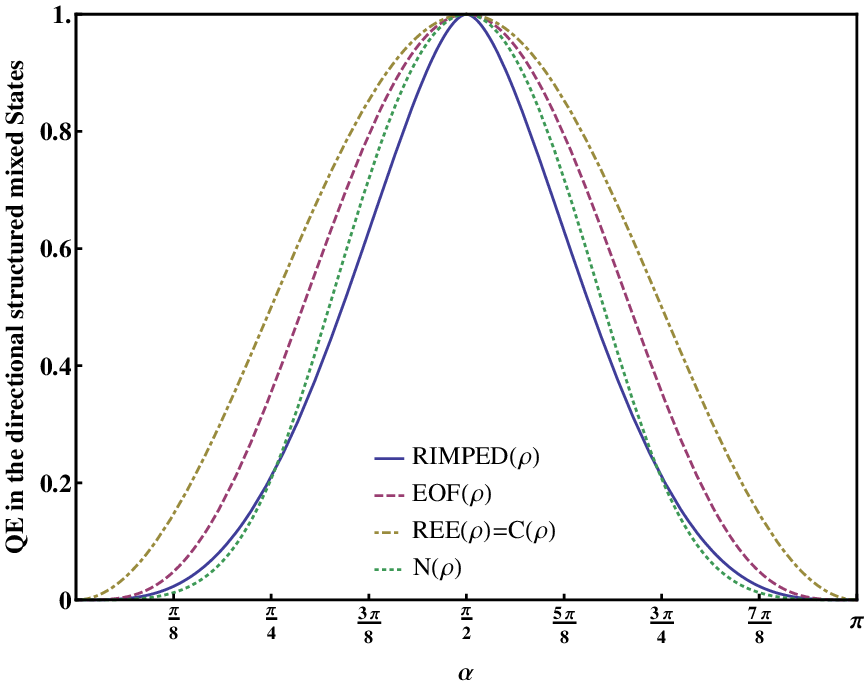}
\renewcommand{\figurename}{Fig.}
\caption{Reduction-induced Minimal Partial Entropy Difference as
well as other some established QE measures 
varying with $\alpha$ in the directional structured mixed states.
When $\alpha=\frac \pi 2$,
$\RIMPED(\rho)=\EOF(\rho)=\REE(\rho)=\C(\rho)=\N(\rho)=1$; whereas
for $\alpha=0, \pi$,
$\RIMPED(\rho)=\EOF(\rho)=\REE(\rho)=\C(\rho)=\N(\rho)=0$. For
values $\alpha=0.8, 2.341$,
$\RIMPED(\rho)=\N(\rho)=0.222676<\EOF(\rho)=0.371412<\REE(\rho)=\C(\rho)=0.515241$;
furthermore, in the ranges $0<\alpha<0.8$ and $2.341<\alpha<\pi$,
$\N(\rho)<\RIMPED(\rho)<\EOF(\rho)<\REE(\rho)=\C(\rho)$; in
contrast, while within the ranges $0.8<\alpha<\frac\pi 2$ and
$\frac\pi 2<\alpha<2.341$,
$\RIMPED(\rho)<\N(\rho)<\EOF(\rho)<\REE(\rho)=\C(\rho)$.}
\end{center}
\end{figure}

As a final example we consider the two qubit Werner state, a highly
symmetric model with global noise
\begin{eqnarray}
\rho_W(p) = p|\Psi^-\rangle\langle\Psi^-| + (1-p)\frac{I}{4}, \qquad
0 \le p \le 1,
\end{eqnarray}
with $|\Psi^-\rangle = (|01\rangle - |10\rangle)/\sqrt{2}$. 
For the symmetric global mixed two qubit Werner state, the form that
reflects the Reduction-induced Minimal Partial Entropy Difference
through its symmetric decomposition into a Bell state and a
maximally mixed separable component, can be alternatively expressed
as
\begin{eqnarray}
\rho_W(p) =
\begin{cases}
(1-3p)\frac I 4 +3p \rho_W(\frac 1 3)  & 0 \le p \le \dfrac{1}{3} \\[10pt]
\displaystyle \frac{3p-1} 2 |\Psi^-\rangle\langle\Psi^-|
+\frac{3-3p} 2 \rho_W(\frac 1 3) & \dfrac{1}{3} < p \le 1
\end{cases}
\end{eqnarray}
where $\rho_W(\frac 1 3)$ for the Werner state at parameter
$p=\frac13$, an explicit separable decomposition can be constructed
as
\begin{eqnarray}
\rho_W\!\left(\frac13\right)=\frac14\sum_{k=1}^4\big|\psi_k\big\rangle\langle\psi_k|\otimes\big|\psi_k\big\rangle\langle\psi_k|
\end{eqnarray}
with four pure single qubit states pointing toward the vertices of a
regular tetrahedron inscribed within the Bloch sphere
\begin{eqnarray}
\begin{aligned}
|\psi_1\rangle&=\sqrt{\frac{3+\sqrt{3}}{6}}|0\rangle+\sqrt{\frac{3-\sqrt{3}}{6}}e^{i\pi/4}|1\rangle,\\
|\psi_2\rangle&=\sqrt{\frac{3-\sqrt{3}}{6}}|0\rangle+\sqrt{\frac{3+\sqrt{3}}{6}}e^{-i\pi/4}|1\rangle,\\
|\psi_3\rangle&=\sqrt{\frac{3-\sqrt{3}}{6}}|0\rangle+\sqrt{\frac{3+\sqrt{3}}{6}}e^{i3\pi/4}|1\rangle,\\
|\psi_4\rangle&=\sqrt{\frac{3+\sqrt{3}}{6}}|0\rangle+\sqrt{\frac{3-\sqrt{3}}{6}}e^{-i3\pi/4}|1\rangle.
\end{aligned}
\end{eqnarray}
From the above decomposition, the entanglement property of the
Werner state can be clearly identified. Specifically, for the
parameter range $0 \le p \le \frac{1}{3}$, the Werner state is fully
separable and possesses no QE. In contrast, when $\frac 1 3 < p \le
1$, the state becomes inseparable and exhibits genuine bipartite QE.
The analytical expressions for the five QE measures introduced above
are as follows.
The Reduction-induced Minimal Partial Entropy Difference (RIMPED),
which coincides with the Relative Entropy of Entanglement (REE), is
computed as
\begin{eqnarray}
&&\RIMPED(\rho_W(p))=\REE(\rho_W(p))
=\nonumber\\&&1+\frac{1+3p}4\log_{2}{\frac{1+3p}4}+\frac{3-3p}4\log_{2}{\frac{3-3p}4}.
\end{eqnarray}
In addition, the Entanglement of Formation (EOF) is expressed as
\begin{eqnarray}
&&\EOF(\rho_W(p))=-\frac{1-\sqrt{1-(\frac{3p-1}{2})^2}}{2}
\log_{2}{\frac
{1-\sqrt{1-(\frac{3p-1}{2})^2}} 2}\nonumber\\
&&-\frac {1+\sqrt{1-(\frac{3p-1}{2})^2}} 2 \log_{2}{\frac
{1+\sqrt{1-(\frac{3p-1}{2})^2}} 2}.
\end{eqnarray}
Moreover, the Concurrence, which is equivalent to the Negativity for the Werner state, is given by
\begin{eqnarray}
\C(p)=\N(p)=\max(0,\frac{3p-1}{2})
\end{eqnarray}
which also represents the probability weight for the entangled
Bell-state component in this statistical ensemble Eq.(26). 
Fig.3 illustrates the behavior of these five bipartite QE measures.
For $0 \le p \le \dfrac{1}{3}$, the measures satisfy
$\RIMPED(\rho)=\EOF(\rho)=\REE(\rho)=\C(\rho)=\N(\rho)=0$; while for
$p=1$, all attain the maximal value
$\RIMPED(\rho)=\EOF(\rho)=\REE(\rho)=\C(\rho)=\N(\rho)=1$. Within
the intermediate range $\frac 1 3<p<1$, these measures follow the
order $\RIMPED(\rho)=\REE(\rho)<\EOF(\rho)<\C(\rho)=\N(\rho)$.
Notably, Concurrence and Negativity yield identical values, form the
upper bound. 
The Entanglement of Formation occupies an intermediate position. 
Most remarkably, the Reduction-Induced Entropy Difference (RIMPED)
coincides exactly with the Relative Entropy of Entanglement, form a
exact coinciding lower bound, revealing a profound intrinsic
connection between minimal structural entropy variation and
optimization-based
distance.

\begin{figure}
\setlength{\abovecaptionskip}{0.3pt}%
\setlength{\belowcaptionskip}{0.1pt}%
\begin{center}
\includegraphics[width=0.45\textwidth]{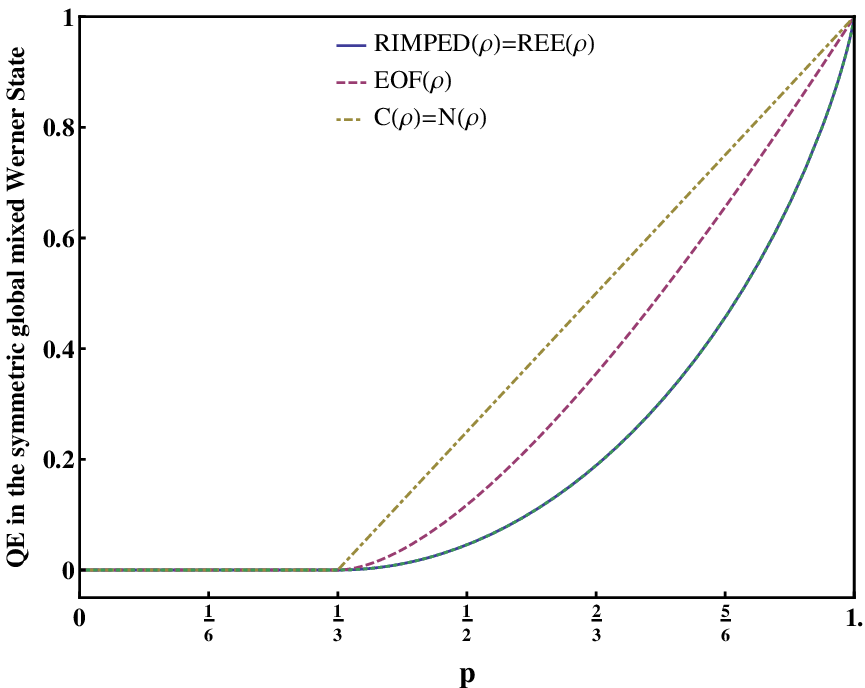}
\renewcommand{\figurename}{Fig.}
\caption{Reduction-induced Minimal Partial Entropy Difference as
well as other some established QE measures in the symmetric global
mixed two-qubit Werner state varying with $p$. In the range $0 \le p
\le \dfrac{1}{3}$,
$\RIMPED(\rho)=\EOF(\rho)=\REE(\rho)=\C(\rho)=\N(\rho)=0$; when
$p=1$, $\RIMPED(\rho)=\EOF(\rho)=\REE(\rho)=\C(\rho)=\N(\rho)=1$;
while within the range $\frac 1 3<p<1$,
$\RIMPED(\rho)=\REE(\rho)<\EOF(\rho)<\C(\rho)=\N(\rho)$.}
\end{center}
\end{figure}

\section{Conclusion and Outlook}
In this paper, we develop the Reduction-Induced Minimal Partial
Entropy Difference (denoted as RIMPED in the figures) as a
structurally motivated quantification of bipartite mixed-state QE.
Built on entropy variation induced by local reduction operations, local \textquotedblleft native\textquotedblright density matrix, which encodes the full local information in contrast to its local reduced density matrix, this method applies to arbitrary bipartite quantum states. 
Partial Von Neumann Entropy is only a pure-stae special case of this
method, 
verifying the theoretical generality of the proposed measure. Three
typical examples covering various structural
complexities, 
show that this measure consistently serves as a conservative lower
bound of QE quantification 
and provides an intrinsic structured perspective,
establishing itself as a valuable complementary tool to existing
measures for quantifying QE in
bipartite mixed states. 
Its intuitive and clear physical picture, 
combined with wide applicability and relatively low computational
complexity, 
indicates the ease of exploring its specific potential
applications in quantum information
science. 
In future studies, the conceptual framework and line of thought
behind the Reduction-Induced Minimal Partial Entropy Difference
deserve to be further developed to describe and quantify
multipartite QE.

\end{CJK}

\begin{references}
\vspace{3mm}

\bibitem{epr} A. Einstein, B. Podolsky, and N. Rosen, Phys. Rev. 47, 777 (1935).
\bibitem{sd} E. Schr\"{o}dinger, Die Naturwissenschaften 23, 807(1935).
\bibitem{bell} J. S. Bell, Physics 1, 195 (1964), reprinted in J. Bell,
Speakable and Unspeakable in Quantum Mechanics, Cambridge University
Press, (2004).
\bibitem{kk} A. K. Ekert, Phys. Rev. Lett. 67, 661 (1991).
\bibitem{ng} N. Gisin, G. Ribordy, W. Tittel, and H. Zbinden, Rev. Mod. Phys. 74,
145 (2002).

\bibitem{ch} C. H. Bennett, G. Brassard, C. Crepeau, R. Jozsa, A.Peres, and W. K. Wootters, Phys. Rev. Lett. 70, 1895 (1993).
\bibitem{rc} Richard Cleve and Harry Buhrman, Phys. Rev. A 56, 1201 (1997).
\bibitem{nr} N Gigena and R Rossignoli, Phys. Rev. A 95, 062320 (2017).

\bibitem{sj} C. H. Bennett and S. J. Wiesner, Phys. Rev. Lett. 69, 2881 (1992).
\bibitem{rr} R. Raussendorf and H. J. Briegel, Phys. Rev. Lett. 86, 5188 (2001).
\bibitem{fa} Frank Arute, Kunal Arya, Ryan Babbush, Dave Bacon, et al., Nature
574, 505 (2019). 

\bibitem{tc} Thomas Chalopin, Chayma Bouazza, Alexandre Evrard, Vasiliy
Makhalov, Davide Dreon, Jean Dalibard, Leonid A Sidorenkov, and
Sylvain Nascimbene, Nature communications 9, 1 (2018).
\bibitem{rcp} R. C. Pooser, N. Savino, E. Batson, J. L.
Beckey, J. Garcia, and B. J. Lawrie, Phys. Rev. Lett. 124, 230504
(2020).
\bibitem{lp} Luca Pezz, Nature Photonics 15,
74 (2021).


\bibitem{ch1} C. H. Bennett, H. J. Bernstein, S. Popesu and B. Schumacher, Phys. Rev. A 53, 2046 (1996); S.Popesu, D.Rohrlich, Phys.
Rev. A 56, R3319(1997).
\bibitem{ch2} C. H. Bennett, D. P. DiVincenzo, J. A. Smolin and W. K. Wootters, Phys. Rev. A 54, 3824 (1996).

\bibitem{wk} S. Hill
and W. K. Wootters, Phys. Rev. Lett. 78(26), 5022(1997); William K. Wootters, Phys. Rev. Lett. 80, 2245 (1998)).

\bibitem{au} A. Uhlmann, Phys. Rev. A 62, 032307 (2000).
\bibitem{ka} K. Audenaert, F. Verstraete and Bart De Moor, Phys. Rev. A 64, 052304
(2001).
\bibitem{mf} S. M. Fei, J. Jost, X. Q. Li Jost, G. F. Wang, Phys. Lett. A
310, 333, (2003); S. M. Fei, X. Q. Li Jost, Rep. Math. Phys. 53, 195
(2004).
\bibitem{xjw} J. W. Xu and Q. H. Chen, Chin. Phys. B Vol.21, No. 4 040302(2012).
\bibitem{pr} P. Rungta, V. Buzek, C.M. Caves, M. Hillery, and G.J. Milburn, Phys.
Rev. A 64, 042315 (2001); S. Albeverio and S. M. Fei, J. Opt. B:
Quantum Semiclass. Opt. 3, 223 (2001).
\bibitem{kc} K. Chen, S. Albeverio, S. M. Fei, 
Phys. Rev. Lett. 95, 040504 (2005).
\bibitem{cs} C.S. Yu, and H.S. Song, 
Phys. Rev. A 80(2), 022324 (2009).
\bibitem{cj} C.J. Zhang, Y.X. Gong, Y.S. Zhang, and G.C. Guo, 
Phys. Rev. A 78(4), 042308 (2008).

\bibitem{zb} Zhi-Bo Chen, Shao-Ming Fei Laser Phys. 35 045205 (2025)






\bibitem{vv1} V. Vedral, M. B. Plenio, M. A. Rip pin, and P. L. Knight, Phys. Rev. Lett. 78, 2275 (1997).
\bibitem{vv2} V. Vedral, M. B. Plenio, Phys. Rev. A 57, 1619 (1998).

\bibitem{K.Z} K. Zyczkowski et al, Phys Rev. A {\bf58} 883 (1998).

\bibitem{gv} G. Vidal and R. F. Werner, Phys. Rev. A 65, 032314, 6899 (2002).

\bibitem{ls} Lewenstein and Sanpera, Phys. Rev. Lett. 80, 2261 (1998).

\bibitem{G.V} G. Vidal, J. Mod. Opt. 47 355 (2000).
\bibitem{Y.C} Y. C. Ou and H. Fan, Phys. Rev. A. 75, 062308 (2007).
\bibitem{W.J} W. D\"{u}r, G. Vidal, J. I. Cirac, Phys. Rev. A 62, 062314(2000).
\end{references}
\end{document}